\documentclass[11pt]{article}

\usepackage{a4wide}
\usepackage{amsmath}
\usepackage{amssymb}
\usepackage{graphicx}
\usepackage{color}

\begin{document}

\title{Smoothing out Negative Tension Brane}
\author{
	Kin-ya Oda$^*$, 
	Takao Suyama$^\dagger$ and
	Naoto Yokoi$^\ddag$\bigskip\\
	$^*$\textit{\small Physics Department, Osaka University, Osaka 564-0063, Japan}\\
	{\small E-mail: \tt
		odakin@phys.sci.osaka-u.ac.jp}\smallskip\\
	$^\dagger$\textit{\small 
%Department of Physics, Kyoto University, Kyoto 606-8502, Japan
Center for Theoretical Physics, Seoul National University, Seoul 151-747, Korea}\\
	{\small E-mail: \tt 
%		suyama@gauge.scphys.kyoto-u.ac.jp
suyama@phya.snu.ac.kr}\smallskip\\
       $^\ddag$\textit{\small Institute of Physics, University of Tokyo, Tokyo 153-8902, Japan}\\
	{\small E-mail: \tt 
		nyokoi@hep1.c.u-tokyo.ac.jp}\bigskip\\
}
\date{March 6, 2009}
\maketitle

\bigskip

\abstract{\noindent
We propose an extension of the five dimensional gravitational action with an external source in order to allow arbitrary smoothing of the negative tension brane in the Randall-Sundrum model.
This extended action can be derived from a model with an auxiliary four form field coupled to the gravity.
%We further show that the model can be viewed as emerging from the tachyon condensation.
We point out a further generalization of our model in relation to tachyon condensation. 
A possible mechanism for radion stabilization in our  model is also discussed.
}

\vfill

\noindent
OU-HET-614/2008\\
SNUTP08-010\\
UT-Komaba/08-19\\
RIKEN-TH-100

\newpage

\normalsize

\section{Introduction}
The huge hierarchy between the weak and Planck scales has been one of the most important issues in the theory of elementary particles.
The braneworld with the warped compactification, which is proposed by Randall and Sundrum, is an attractive model to explain the hierarchy between the weak and Planck scales~\cite{Randall:1999ee}.
%Although 
Even though the hierarchy problem is resolved in the Randall-Sundrum scenario, 
we still need a fine tuning among the bulk cosmological constant and 
the brane tensions in order to acquire the observed four-dimensional cosmological constant 
$\Lambda_4\lesssim 10^{-120}M_P^4$, where $M_P$ is the four-dimensional Planck scale 
determined by the Newton's law of gravitation in four dimensions.
%\begin{itemize}
%	\item In particular, this fine tuning requires the infrared (IR) brane tension to be negative.
%A negative tension brane is known to be problematic in general relativity
%~\cite{Shiromizu:1999wj,Ida:2001qw,Ochiai:2001fh,Maity:2006ub}, 
%see also~\cite{Csaki:1999mp,Maity:2006in} for possible resolution.
%	\item Actually, above logic turned out to be out of date; I asked Shiromizu-san 
%and he says there's no problem for classical gravity if we fix radion; 
%Only possible flaw is the semi-classical bubble formation~\cite{Ida:2001qw,Ochiai:2001fh}.
%	\end{itemize}
This fine tuning requires that one of the branes should have a negative tension.
Although the negative tension brane appears to lead to some pathology in general relativity~\cite{Shiromizu:1999wj} (see also~\cite{Aliev:2004ds}), detailed analyses have revealed that the four-dimensional effective theory on the negative tension brane can be consistent with the standard Einstein gravity~\cite{Kanno:2002ia,Shiromizu:2002qr} as far as the moduli for the compactification radius, the radion, is fixed by some mechanism, such as the one proposed by Goldberger and Wise~\cite{Goldberger:1999uk}.

%Another aspect of the hierarchy problem is the stability of the weak scale in the radiative corrections.
%Supersymmetry is one of the solutions for this stability problem.
Recent progress in string theory has revealed 
that the compactified space is generically warped under the presence of 
background fluxes~\cite{Giddings:2001yu}, and 
much effort to realize Randall-Sundrum scenario in superstring theory has been made, see e.g.~\cite{Brummer:2005sh} and references therein.
The realization of Randall-Sundrum scenario in superstring theory 
naturally requires the supersymmetric extension of the Randall-Sundrum model in five-dimensional
effective supergravity.  

So far, two ways to supersymmetrize the Randall-Sundrum model are proposed in~\cite{Gherghetta:2000qt,Falkowski:2000er} and in~\cite{Altendorfer:2000rr}, respectively.
The former involves a kinky gauge coupling which has position dependence 
like a step function in the extra dimension.
Especially the multiple of the step function and its derivative, the delta function, vanishes everywhere.
The latter does not involve the position dependent gauge coupling 
but it implicitly assumes that the multiple of a step function 
and a delta function takes a finite non-zero value on the branes~\cite{Bergshoeff:2000zn,Fujita:2001bd}.\footnote{We thank Y.~Sakamura for pointing out this issue.}

We want to have a way to regularize, {\it i.e.} smooth out, the step function and the delta function in the Randall-Sundrum geometry in particular on the brane.
The step function would be realized as a kink solution of a bulk field.
The positive tension brane has been smoothed out 
by introducing a bulk scalar field with solitonic kink solutions~\cite{DeWolfe:1999cp}.
It has been shown that the negative tension brane cannot be smoothed by a bulk scalar field without an instability of the spacetime~\cite{Koley:2004au,Pospelov:2004aw,Nunes:2005up,Koley:2005nv}.

Here we propose an extension of the five-dimensional Einstein-Hilbert action 
that allows the smoothing of the negative tension brane.
%In our regularized model, the multiple of the step function and 
%the delta function takes a different value in the singular limit 
%from those in~\cite{Bergshoeff:2000zn,Fujita:2001bd}, 
%so our model might be able to provide a new way to supersymmetrize the Randall-Sundrum model.
In our regularized model, the multiple of the step function and the delta function vanishes 
on the branes in the singular limit, 
%takes a different value from those in~\cite{Altendorfer:2000rr} and 
in agreement with~\cite{Gherghetta:2000qt,Falkowski:2000er}.
Our model might provide an insight on the supersymmetrization of the Randall-Sundrum model without resorting to singular configurations.

Besides the somewhat technical motivation presented above, our study can be placed on more general physical ground. 
Singular objects such as D-branes in string theory are often interpreted as 
solitonic smooth objects in dynamical or quantum mechanical analysis.
Our regularization could be a useful tool for these analyses in the braneworld models.
%One would expect that a singular object is generically smoothed 
%when one takes dynamical or quantum effects into account. 
%Our regularization could be a useful tool to understand these phenomena in brane-world models.

In the next section, we describe our model of the extension of 
five-dimensional Einstein gravity.
In Section~3, we present a possible embedding of our model into the standard five-dimensional gravity 
with a four-form field. Further extension with a coupling between the four-form and a scalar source 
is also shown, inspired by the tachyon condensation in string theory.
In Section~4, we show that the radion can be stabilized by a natural extension of the Goldberger-Wise mechanism.
In the last section, we provide summary and discussions.

\section{Modified Gravitational Action} 
Let us first briefly review the Randall-Sundrum model~\cite{Randall:1999ee}.
The fifth dimension $y$ is compactified on $S^1/Z_2$ by the identifications
	$y\sim y+2 L$ and 
	$-y\sim y$.
When we restrict it to $-L<y\leq L$, the gravitational part of the action reads
\begin{align}
 S_0 &= 
%  {M_*^3}\int d^5x\left[\sqrt{-g}\left({R\over2}-\Lambda\right)-\sqrt{-g^{(4)}}\bigg(\lambda_0\delta(y)+\lambda_L\delta(y-L)\bigg)\right],
 {M_*^3}\int d^5x\sqrt{-g}\left({R\over2}-\Lambda-\lambda_0\delta(y)-\lambda_L\delta(y-L)\right),
 \label{original_action}
 \end{align}
where
	%$g^{(4)}$ is the four dimensional induced metric and
	$M_*^3=1/8\pi G$ with $G$ being the higher dimensional Newton constant.\footnote{When we expand around the flat background $g_{MN}=\eta_{MN}+h_{MN}$, the graviton $h_{MN}$ is canonically normalized with the unit $32\pi G=1$.} 
%, and $R$ is the Ricci scalar in five dimensions.
Note that this action does not have the full five-dimensional diffeomorphism invariance, but the only four-dimensional diffeomorphism invariance and re\-pa\-ram\-e\-tri\-za\-tion invariance with respect to the compact direction $y$.

The Randall-Sundrum geometry is a slice of the five dimensional anti de Sitter space (AdS$_5$):
\begin{align}
 ds^2 &= e^{-2k|y|}\eta_{\mu\nu}dx^\mu dx^\nu+dy^2,
 \label{metric}
 \end{align}
where $k$ is its curvature scale.
The metric~\eqref{metric} becomes a solution to the five dimensional Einstein equation $\delta S/\delta g^{MN}=0$
if the five dimensional bulk cosmological constant $\Lambda$ and the brane tensions $\lambda_0$ and $\lambda_L$ are fine-tuned to be
\begin{align}
 \Lambda=-k\lambda_0=k\lambda_L=-6k^2<0.
 \label{CC_fine_tuning}
 \end{align}
Physically the condition~\eqref{CC_fine_tuning} implies that the ultraviolet (UV) brane at $y=0$ has the positive tension while the infrared (IR) brane at $y=L$ has the negative one.
In~\cite{DeWolfe:1999cp} it has been shown that the positive tension brane can be 
smoothed out by introducing a bulk scalar having a kink profile.

There have been attempts to smooth out the negative tension brane by introducing a so-called ghost scalar, which is a propagating degree of freedom having a wrong-sign kinetic term~\cite{Koley:2004au, Pospelov:2004aw, Nunes:2005up}.
Obviously there are resulting difficulties both at the classical and quantum levels. 
Here we attempt to smooth out the negative tension brane in a different way.
A possible resolution to the wrong sing kinetic term is shown in the next section.

Now we propose a modification of the action~\eqref{original_action} preserving the same symmetry, that is the four dimensional diffeomorphisms and the re\-pa\-ram\-e\-tri\-za\-tion of the fifth coordinate $y$:
\begin{align}
 S_g &= M_*^3\int d^5x\sqrt{-g}\left({R\over2}-\Lambda\varepsilon(y)^2-\lambda{\varepsilon'(y)\over2\sqrt{g_{yy}}}\right),
 \label{new_action}
 \end{align}
where $\varepsilon(y)$ is an arbitrary $Z_2$ odd periodic function:
\begin{align}
 \varepsilon(y+2L) &= \varepsilon(y), &
 \varepsilon(-y)   &= -\varepsilon(y),
 \label{parity}
 \end{align}
and $\varepsilon'(y)=\partial_{y}\varepsilon(y)$.

If we again fine tune the cosmological constant and the brane tension:
\begin{align}
	\Lambda=-k\lambda=-6k^2<0,
	\label{new_fine_tuning}
 \end{align}
%the solution to the Einstein equation
the Einstein equation
\begin{align}
%	{\delta S\over\delta g^{MN}}\propto
		R_{MN}-{g_{MN}\over2}R
		+\Lambda\varepsilon(y)^2g_{MN}
   	+\lambda{\varepsilon'(y)\over2\sqrt{g_{yy}}}
		\left(g_{MN}-{g_{yM}g_{yN}\over g_{yy}}\right)
			&=0
 \end{align}
%becomes the metric:
has a solution:
\begin{align}
 ds^2=e^{-2\sigma(y)}\eta_{\mu\nu}dx^\mu dx^\nu+dy^2,\label{sigma_metric}
 \end{align}
where
\begin{align}
 \sigma(y) &= k\int^{y} d\tilde{y}\,\varepsilon(\tilde{y}).
 	\label{sigma_epsilon}
 \end{align}
We note that the equation of motion of $\varepsilon(y)$:
\begin{equation}
2\Lambda\varepsilon(y) 
= \frac{\lambda}{2\sqrt{-g}}\partial_y\left(\frac{\sqrt{-g}}{\sqrt{g_{yy}}}\right), 
\end{equation}
is automatically satisfied due to Eqs.~(\ref{new_fine_tuning}) and (\ref{sigma_epsilon}).

Now we can have an arbitrary shape of the $Z_2$ even periodic function $\sigma(y)$ by a proper choice of $\varepsilon(y)$.
A choice of the external source $\varepsilon(y)$ gives a corresponding gravitational action.
Any choice of $\varepsilon(y)$ can satisfy the Einstein equation as well as the equation of motion of $\varepsilon(y)$ itself, with the tuning~\eqref{new_fine_tuning}.
One might wonder what is the source of such a large symmetry.
We come back to this point in the next section.

\begin{figure}
	\begin{center}
		\mbox{\includegraphics[width=.3\linewidth]{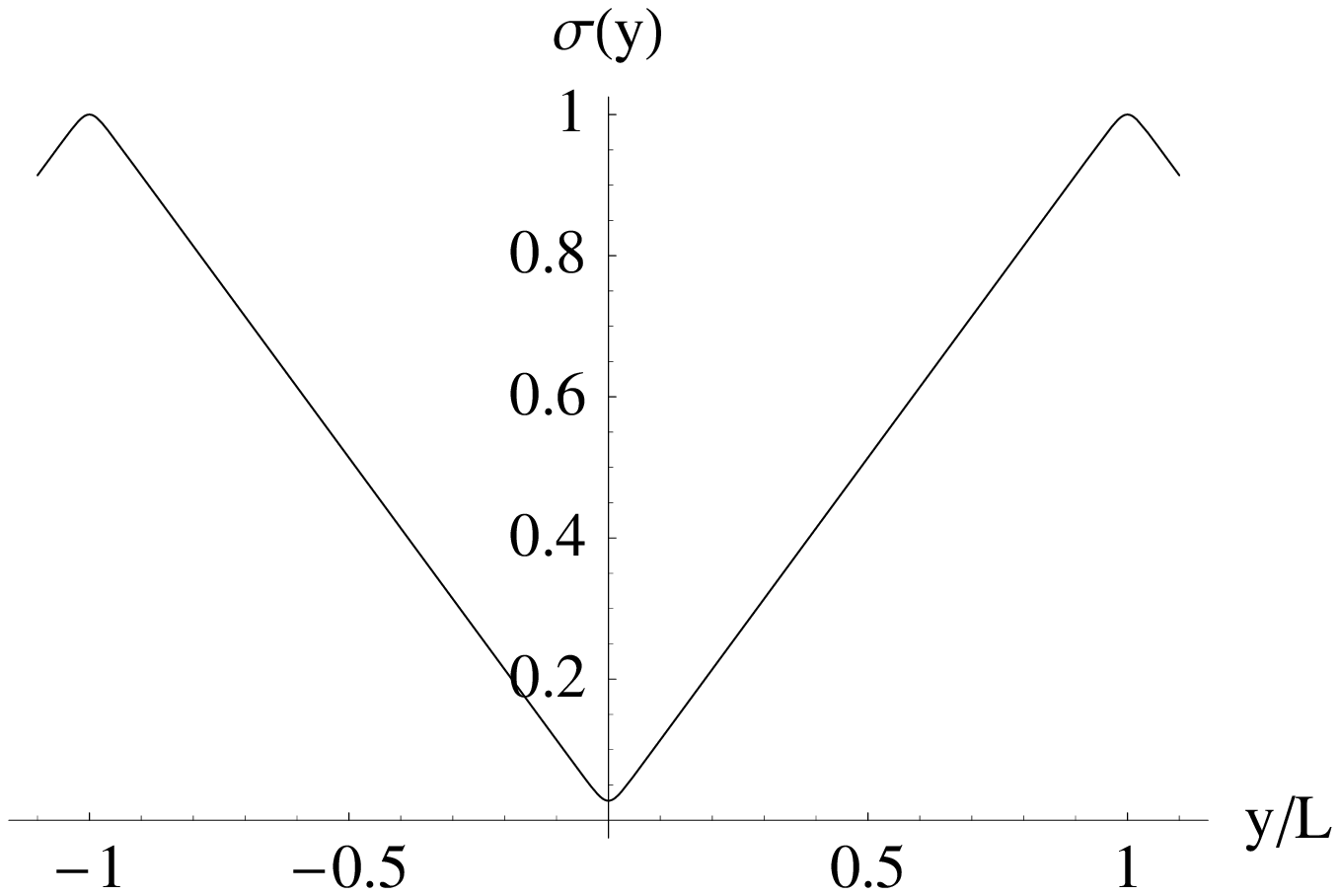}}
		\hfill
		\mbox{\includegraphics[width=.3\linewidth]{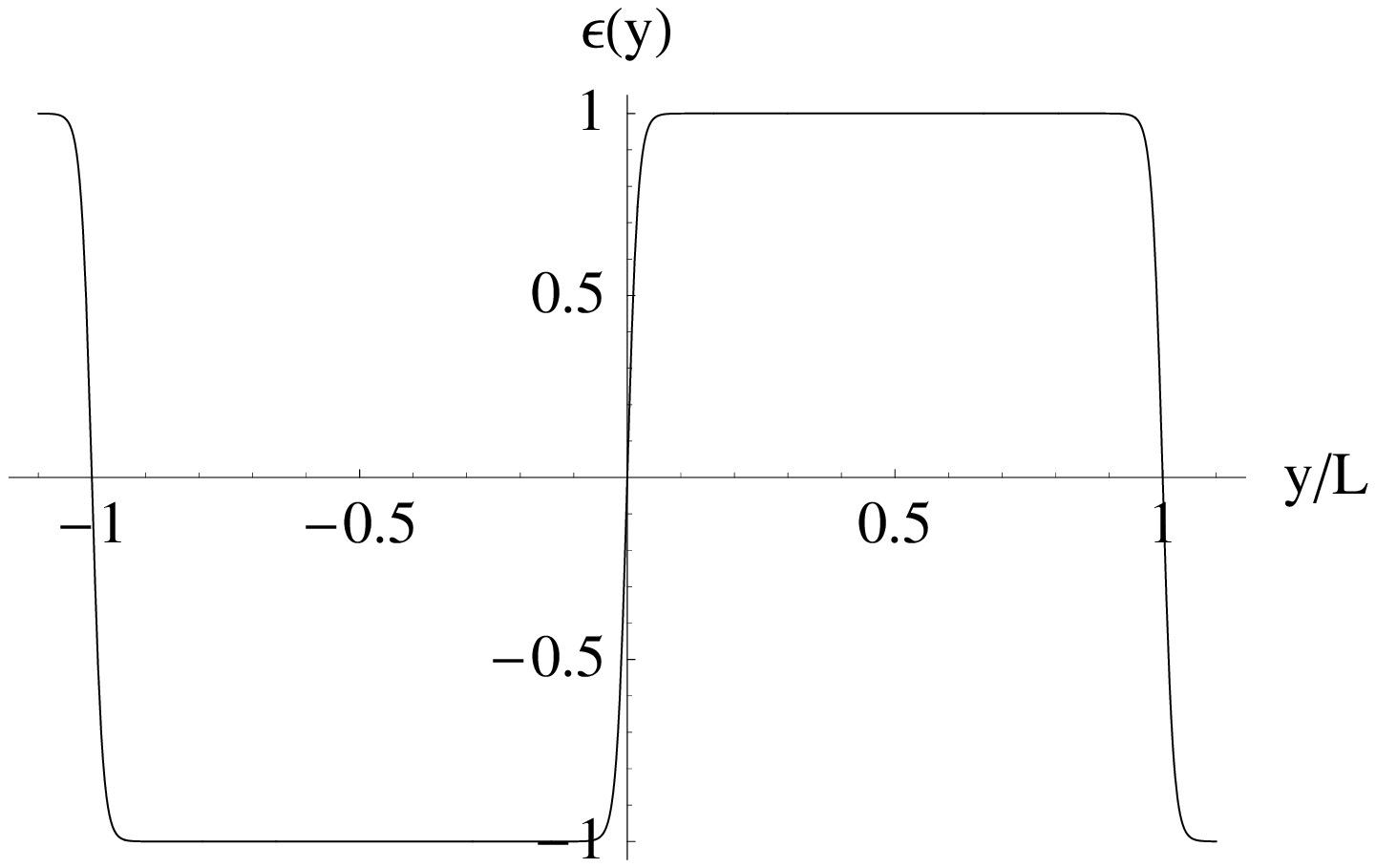}}
		\hfill
		\mbox{\includegraphics[width=.3\linewidth]{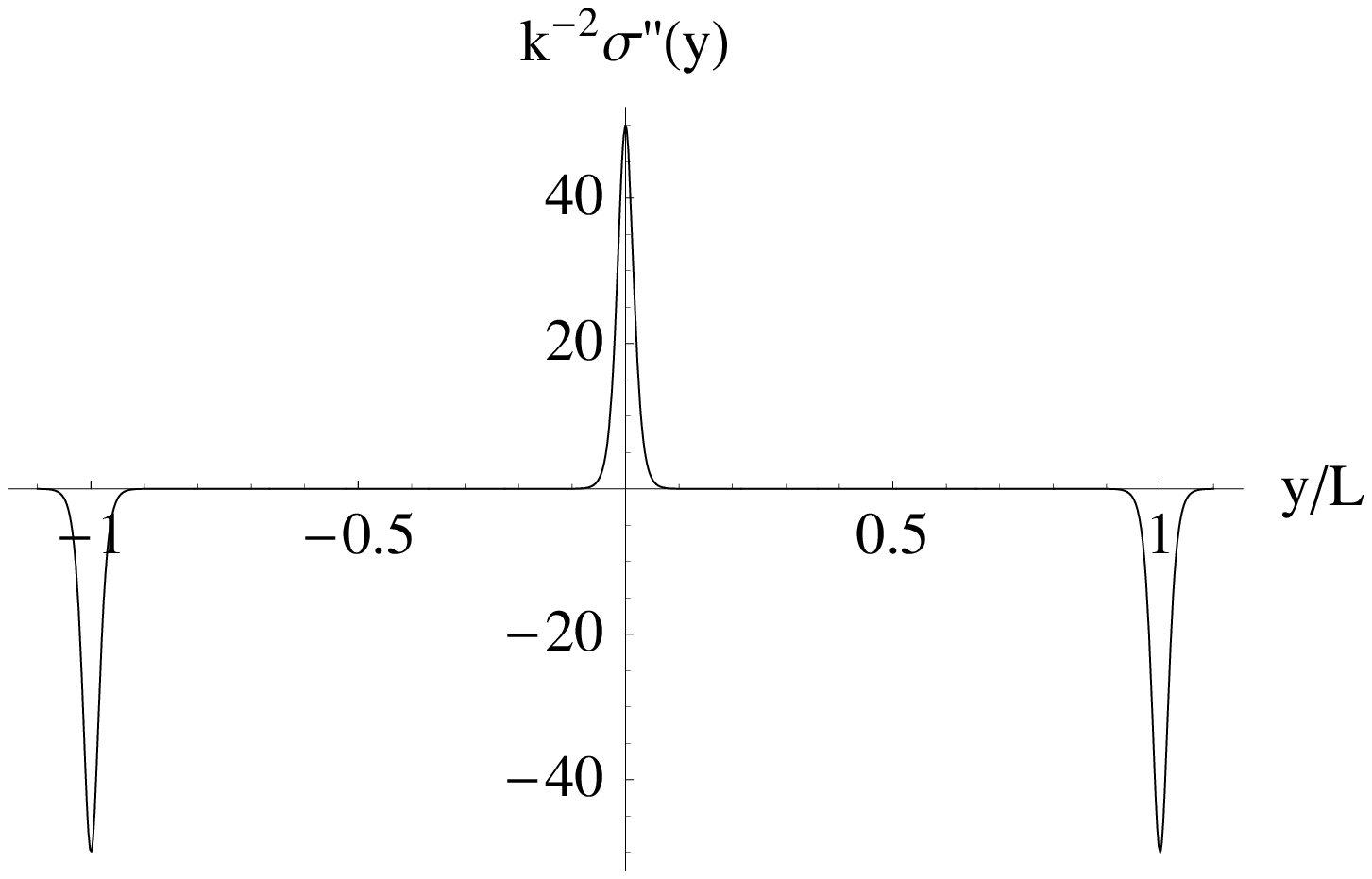}}
		\caption{Smoothed functions $\sigma(y)$ (left), $\varepsilon(y)=\sigma'(y)/k$ (center), and $\varepsilon'(y)$ (right) for $\beta=50$.}
		\label{smoothed_funcs}
	  \end{center}
 \end{figure}

When we take $\varepsilon(y)$ to be the step function:
\begin{align}
	\varepsilon(y) &= \begin{cases}1&(0<y<L),\\ -1&(-L<y<0),\end{cases}
	\label{step_func}
 \end{align}
we recover the original setup~\eqref{original_action} with $\lambda=\lambda_0=-\lambda_L$.
The point here is that a continuous deformation of the negative tension brane becomes possible.
For instance, we can take\footnote{See e.g.\ footnote~5 in Ref.~\cite{Oda:2004rm}.}
\begin{align}
 \varepsilon(y) &= -\tanh[\beta(y+L)]+\tanh\beta y-\tanh[\beta(y-L)]
	\label{smooth_eps}
 \end{align}
for $-L\leq y\leq L$.\footnote{It is straightforward to make~\eqref{smooth_eps} symmetric under the translation $y\rightarrow y+2L$ by adding terms at other integer multiples of $L$.}
In the $\beta\to\infty$ limit, $\varepsilon(y)$ in~\eqref{smooth_eps} goes back to the step function~\eqref{step_func}.
For~\eqref{smooth_eps}, the $\sigma(y)$ is obtained as\footnote{The value of the constant term can 
be varied by rescaling $x^\mu$.} 
%?????I?? rescale ???????????????????B?i???????P???????????????????????????B?j
\begin{align}
	\sigma(y) &= {k\over\beta}\log{\cosh\beta y\over\cosh[\beta(y-L)]\cosh[\beta(y+L)]}+2.
       \label{smooth profile}
 \end{align}
%for the fine tuning condition~\eqref{new_fine_tuning}.
As an example, the plot for $\beta=50$ is shown in Fig.~\ref{smoothed_funcs}. Note that this gives 
a smooth regularization of the Randall-Sundrum geometry which is fully consistent with the bulk 
Einstein equation. 

In our formulation, the multiple of the (regularized) step and delta functions is always zero 
on the branes,
\begin{align}
 \varepsilon(y)\varepsilon'(y)=0 \qquad \text{at $y = 0, L$}.
 \end{align}
Therefore, it may be utilized as a tool to supersymmetrize the Randall-Sundrum model in the formulation of~\cite{Gherghetta:2000qt,Falkowski:2000er}, but not directly for~\cite{Altendorfer:2000rr}.

\section{Four Form Field and Tachyon Condensation}
So far, the action~\eqref{new_action} is given just by hand.
One might consider that this is too ad hoc.
Here we argue that the action~\eqref{new_action} could be regarded as an effective theory of an underlying microscopic theory. 

Let us consider the following five dimensional gravitational action coupled to a four-form gauge field $A_4$: 
\begin{align}
 S &= {M_*^3\over2}\int\left(R\wedge*1+F_5\wedge*F_5\right),
 \label{4-form_action}
 \end{align}
where $F_5=dA_4$ and $*$ is the Hodge dual in five dimensions.
Note that the gauge kinetic term has a wrong sign here.
This does not lead to an immediate inconsistency because $A_4$ has no physical degree of freedom in five dimensions.
Such a wrong-sign ``auxiliary'' gauge field has also been introduced by Turok and Hawking for an inflation model inspired by the M-theory~\cite{Turok:1998he}.

Generically one can write a five form field in five dimensions as
\begin{align}
 F_5  &= f(x)\sqrt{-g}\,d^5x, &
 *F_5 &= -f(x),
 \end{align}
where $d^5x=dx^0\wedge\dots\wedge dx^3\wedge dy$ and $f(x)$ is a scalar function.
The action in terms of $f(x)$ reads
\begin{align}
	  S &= {M_*^3\over2}\int \sqrt{-g}\,d^5x\left(R-f^2\right)= 
{M_*^3\over2}\int \sqrt{-g}\,d^5x\left(R+\varphi^2-2\varphi f\right),
 \end{align}
where we have introduced another auxiliary field $\varphi$ in the last step.

The gauge field $A_4$ has five independent component fields. 
Assuming the invariance under the four dimensional diffeomorphisms and 
the re\-pa\-ram\-e\-tri\-za\-tion of~$y$, it would be natural to expect that 
the components of $A_4$ except for $A_{\mu_1\mu_2\mu_3\mu_4}$, where $\mu_i=0,\cdots,3$, are 
irrelevant, and therefore they could be ignored from the beginning. 
%Assuming the invariance under the four dimensional diffeomorphisms and 
%the re\-pa\-ram\-e\-tri\-za\-tion of $y$, the four form gauge field generally takes the following form
For the remaining component, we make the following field redefinition: 
\begin{align}
	A_4 &= a(x)\sqrt{-g^{(4)}}\,d^4x,
\label{4-form_ansatz}
 \end{align}
where $a(x)$ is a scalar function and $d^4x=dx^0\wedge\dots\wedge dx^3$.
In terms of $a(x)$, the field strength becomes
\begin{align}
 f &= {1\over\sqrt{-g}}\partial_y\left(a(x)\sqrt{-g^{(4)}}\right).
 \end{align}
Putting this into the action, we obtain 
\begin{align}
	S &= {M_*^3\over2}\int d^5x\sqrt{-g}\left(R+\varphi^2+{2a(x)\over\sqrt{g_{yy}}}\partial_y\varphi\right),
   \label{rewrite}
 \end{align}
where we have performed the partial integration.
If we could choose $\varphi(x)$ and $a(x)$ as 
\begin{align}
	\varphi(x) &= \sqrt{\Lambda}\,\varepsilon(x), &
	a(x)       &= {\lambda\over4\sqrt{\Lambda}},
 \end{align}
we recover the extended action~\eqref{new_action}.

Now it is obvious why $\varepsilon$ can be an arbitrary function, 
that is, why $\delta S/\delta\varepsilon=0$ is always satisfied
because $\varepsilon$ is nothing but an auxiliary field $\varphi$.
The fine tuning of the cosmological constant and the brane tension 
is done by fixing the constant value of the scalar field $a(x)$.
We note that $a(x)$ must be treated 
as a genuinely external source, as long as the action~\eqref{4-form_action} is regarded as the microscopic action of our theory, because the equation of motion for $a(x)$ leads to
the unwanted condition $\varepsilon'(y)=0$ which does not allow a kink-like solution.

It is interesting to observe that the action~\eqref{4-form_action} can be rewritten as~\eqref{rewrite} which is similar to~\eqref{new_action}.
%It might suggest that 
There might be a more fundamental theory whose action includes the terms in~\eqref{4-form_action} % and possibly more with the other kinds of fields. 
so that our action~\eqref{new_action} can be regarded as (a part of) its effective action. 
Then the resultant equation of motion of $a(x)$ would be more complicated, hopefully 
allowing a non-trivial solution for $\varphi(x)$. 
The constant $a(x)$ solution should be consistent with the equations of motion of the fundamental theory. 

In the more fundamental theory, the profile of $\varepsilon(y)$, which has been completely arbitrary so far, might be determined by some mechanism. 
For example, suppose that there is the following term of the Chern-Simons type in the action of the underlying theory:
\begin{align}
S_{CS} &= %	-\int T(x)\,F_{5} = 
	\int dT \wedge A_{4},
\end{align}
where $T$ is a $Z_{2}$-odd scalar source. 
Due to this term, the equation of motion of $a(x)$ can relate $\varphi(x)$ to $T(x)$. 
As a result, if $T(x)$ has a kink-like profile, so would $\varphi(x)$. 
Note that this term is topological, {\it i.e.} independent of metric 
and does not change the Einstein equation. % of motion from metric variations (?v?`?F?b?N?j.
We also note that this extra term often appears in string theory in the context of the tachyon condensation 
where the $T(x)$ is indeed identified as a tachyon field, see~\cite{Sen:1999mg} for a review 
and references therein.
Further investigation of a possible underlying theory that can lead to our action is left for future research.

%The verification of our expectation that~\eqref{new_action} would be a consistent truncation of an underlying theory requires a complete knowledge, which is beyond the scope of this paper. 

%The justification of our claim that~\eqref{new_action} would be a kind of the consistent truncation of an underlying theory requires the complete knowledge of the underlying theory, which is beyond the scope of this paper. 

%The difficulty is now transferred to the tachyon condensation sector 
%whose study is still being in progress, see e.g.~\cite{Suyama:2007vh} and references therein.

\section{Introducing bulk scalar field}
In general a bulk scalar field plays a crucial role in the radion stabilization~\cite{Goldberger:1999uk,DeWolfe:1999cp}.
To see the dependence on radius $L$ explicitly in our setup, we rewrite our background metric~\eqref{metric} as
\begin{align}
	ds^2	&=	e^{-2kL\alpha(\theta)}\eta_{\mu\nu}dx^\mu dx^\nu+L^2d\theta^2,
 \end{align}
where $\alpha(\theta)=\alpha(\theta+2)$ is an arbitrary dimensionless periodic function.
One can check that the equation of motion for the metric is satisfied for an arbitrary $L$, 
which is promoted to a massless field, radion, in the perturbation analysis.

%The Randall-Sundrum model does not provide a correct four dimensional gravity at the negative-tension brane without the radion stabilization~\cite{Kanno:2002ia,Shiromizu:2002qr}.
%We can show that our modification~\eqref{new_action} can be compatible with the Goldberger-Wise mechanism~\cite{Goldberger:1999uk}.
In our setup having symmetries of four dimensional diffeomorphisms and re\-pa\-ram\-e\-tri\-za\-tion of the fifth coordinate, a bulk scalar field $\Phi$ can couple to the gravity in the following form
\begin{align}
 S
	&= M_*^3\int d^5x\sqrt{-g}\left[{R\over2}-\Lambda(\Phi)\varepsilon^2-\lambda(\Phi){\partial_5\varepsilon\over2\sqrt{g_{55}}}-{g^{MN}\over 2M_*^3}\partial_M\Phi\partial_N\Phi\right],
	\label{radion_action}
 \end{align}
where the cosmological constant $\Lambda$ and the brane tension $\lambda$ are now promoted to 
functions of~$\Phi$. The mass dimensions are: $[\Phi]=3/2$, $[\Lambda]=2$, and $[\lambda]=1$.
Assuming that the vacuum expectation value is much smaller than the Planck scale $\Phi^2\ll M_*^3$, we can expand in terms of $(\Phi^2/M_*^3)$
\begin{align}
	\Lambda(\Phi) &= M_*^2\left(c_1 +{c_2\over2} {\Phi^2\over M_*^3}+{c_3\over4!}\left({\Phi^2\over M_*^3}\right)^2+\cdots\right), \nonumber\\
	\lambda(\Phi) &= M_*  \left(d_1 +{d_2\over2} {\Phi^2\over M_*^3}+{d_3\over4!}\left({\Phi^2\over M_*^3}\right)^2+\cdots\right),
 \end{align}
%we can show that the radion can be stabilized.\footnote{?v?`?F?b?N}
where we have assumed the invariance under the flip $\Phi\rightarrow -\Phi$ for simplicity.

To allow the solution of the form of the generalized step function $\varepsilon(\theta)=\alpha'(\theta)$, we set the first order constants to be $c_1=-6(k/M_*)^2$ and $d_1=6(k/M_*)$.
Note that the fine tuning of the first order constants $c_1$ and $d_1$ corresponds to the fine-tuning of the cosmological constant, which is inevitable for all the versions of Randall-Sundrum type models so far.

The resultant bulk scalar field equation is
\begin{align}
	\Phi''(\theta)
	-4kL\alpha'(\theta)\Phi'(\theta)
	-L^2M_*^3\left[
		\alpha'(\theta)^2{\partial\Lambda\over\partial\Phi}
		+{\alpha''(\theta)\over2L}{\partial\lambda\over\partial\Phi}\right]=0.
		\label{bulk_scalar_eq}
 \end{align}
Once we find the solution to the field equation~\eqref{bulk_scalar_eq} for $\Phi$, 
generically a potential for the radion $L$ in the four-dimensional effective theory 
is generated as
\begin{align}
	V(L)&=	LM_*^3\int_{-1}^1d\theta\,e^{-4kL\alpha(\theta)}\left(
			\alpha'(\theta)^2\Lambda(\Phi)
			+{\alpha''(\theta)\over2L}\lambda(\Phi)
			+{1\over 2L^2M_*^3}\Phi'(\theta)^2
			\right).
 \end{align}

Since, in general,  the equation~\eqref{bulk_scalar_eq} is highly non-linear, 
we cannot expect to have an analytic solution.\footnote{Of course, it is straightforward 
to solve the differential equations numerically.}  
However, for free scalar field, one can obtain an analytic solution for a certain choice of 
parameters as we show below. By an ansatz $\Phi(\theta)=v^{3/2}e^{\sigma(\theta)}$, 
the field equation~\eqref{bulk_scalar_eq} reads
\begin{align}
	\sigma''(\theta)+\sigma'(\theta)^2
	-4kL\,\alpha'(\theta)\sigma'(\theta)
	-\left(LM_*\right)^2\left[c_2+\cdots\right]\alpha'(\theta)^2
	-{LM_*\over2}\left[d_2+\cdots\right]\alpha''(\theta)
		=0,    \label{Riccati}
 \end{align}
where terms denoted by $\cdots$ are $O(v^3/M_*^3)$ and neglected in the following. The parameter $v$ becomes 
the vacuum expectation value of $\Phi$ at the positive tension brane 
in the limit~\eqref{step_func} with the normalization $\alpha(0)=0$.
Here, we set the second order constants $c_2={d_2\over4}\left(d_2-{8k\over M_*}\right)$ 
so that we can have an analytic solution
\begin{align}
	\sigma(\theta)
		&=	{d_2\over2}LM_* \alpha(\theta).  \label{specialsoln}
 \end{align} 
There should be a more general solution to the equation (\ref{bulk_scalar_eq}), 
since it is the second order differential equation while the solution 
specified by (\ref{specialsoln}) has only one integration constant $v$. 
To examine the general solution, it is convenient to define 
$\Psi(\theta)=e^{-d_{2}LM_* \alpha(\theta)/2}\Phi(\theta)$.
The equation for $\Psi(\theta)$ is reduced to
\begin{equation}
\Psi^{''}(\theta)+\left(d_{2}M_* -4 k\right) L \alpha^{'}(\theta)\Psi^{'}(\theta) = 0.
\end{equation}
Obviously, $\Psi(\theta)=\textrm{const.}$ is a trivial solution. The above equation can be 
integrated easily and the general solution becomes
\begin{equation}
\Psi(\theta)=A + B \int^{\theta} d\tilde{\theta} 
\exp\left(\left(d_{2}M_* -4 k\right) L \alpha(\tilde{\theta})\right).
\end{equation}
Since the integral in the second term does not satisfy periodicity condition 
$\Psi(\theta+2)=\Psi(\theta)$, one has to put $B=0$. This proves that
the solution (\ref{specialsoln}) gives the general periodic solution to Eq.~(\ref{bulk_scalar_eq}).

Using this solution, we have the effective potential for the radius $L$ 
up to quartic order ($\sim \Phi^{4}$):  
%becomes
\begin{align}
	V(L)%&=	LM_*^3\int_{-1}^1d\theta\,e^{-4kL\alpha(\theta)}\left(
			%\alpha'(\theta)^2\Lambda(\Phi)
			%+{\alpha''(\theta)\over2L}\lambda(\Phi)
			%+{1\over 2L^2M_*^3}\Phi'(\theta)^2
			%\right)\nonumber\\
%		&=	M_*v^3\int_{-1}^1d\theta\,e^{LM_*(d_2-{4k\over M_*})\alpha(\theta)}\left[
%			{d_2\over4}LM_*\left(d_2-{4k\over M_*}\right)\alpha'(\theta)^2
%			+{d_2\over4}\alpha''(\theta)
%			\right].
		&=	{Lv^6\over24M_*}
					\left(c_3-d_3\tilde d_2\right)
				\int_{-1}^1d\theta\,\alpha'(\theta)^2
			e^{2\tilde d_2 LM_*\alpha(\theta)},
				%{\color{red}
				%	{M_*^2\over8}e^{-2kL\alpha(\theta)}
				%	\left(\tilde d_2+{2k\over M_*}\right)^2
				%}
			\label{radion_potential}
 \end{align}
where we have defined $\tilde d_2\equiv d_2-{2k\over M_*}$. 
Note that $c_1$ and $d_1$ terms are being understood to cancel the gravitational part 
and hence omitted and that the quadratic $\Phi^2$ terms vanish too.
%We have performed partial integration in the last step.
Since all the non-linear terms are suppressed by $v^3/M_*^3$, our free field solution and 
effective potential correspond to the leading approximations in the $v^3/M_*^3$ expansion.  

\begin{figure}
	\begin{center}
		\mbox{\includegraphics[width=.5\linewidth]{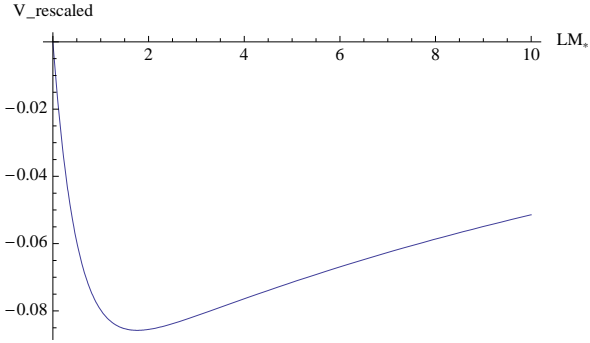}}
		\caption{Sample radion potential being rescaled as $V(LM_*)/{v^6\over 24M_*^2}$ 
vs.\ dimensionless redius $LM_*$ for parameters $\tilde d_2=-1$ and $c_3-d_3\tilde d_2=-0.1$.}
		\label{fig_radion}
	  \end{center}
 \end{figure}

In Fig.~\ref{fig_radion}, we plot the potential $V(L)$ with the regularized profile 
(\ref{smooth profile}) for $\beta=50$ and the parameters  
$\tilde d_2=-1,~c_3-d_3\tilde d_2=-0.1$ as an illustration.
We see that the non-trivial potential for $L$ is generated as the first approximation in the $v^{3}/M_*^{3}$ expansion.
%
%Indeed we can obtain the nontrivial potential for $L$ in this sample analysis, 
%but the potential minimum value is not zero and negative. 
%We have to further fine-tune the cosmological constant by shifting $c_1, d_1, c_2, d_2$.
%
%Note that the negative coefficient for $v^6 \sim  \Phi^{4}$ term in 
%the effective potential does not 
%necessarily 
%lead to the unbounded potential for $\Phi$, because of higher order terms 
%suppressed by $1/M_{*}^{3}$.
Several comments are in order:
\begin{itemize}
\item	The resultant potential minimum value is negative here and we have to fine-tune the cosmological constant again by shifting $c_1, d_1, c_2, d_2$.
\item   In the singular limit, $\alpha(\theta) \rightarrow |\theta|$, the effective potential 
becomes a monotonically decreasing function of $L$, and does not have a minimum. 
The non-trivial minimum is generated due to the smoothness of the branes.
\item	The original Goldberger-Wise setup~\cite{Goldberger:1999uk} corresponds to $c_3=0$, where it has been crucial to have different (positive) brane potentials from each other.\footnote{The different potentials give the different boundary conditions on the bulk scalar field that lead to the non-trivial wave function profile in the extra dimension. Then the kinetic term balances the bulk mass term in the total energy and serves a ``repulsive force'' to keep a finite radius for the extra dimension, see e.g.\ Ref.~\cite{Csaki:2004ay}.}
However, in our action~\eqref{radion_action},
the brane potential is given by a single function $\lambda(\Phi)$ and cannot be independent of each other even in the singular limit. In particular, the sign of quartic coupling in the brane potential 
is flipped by the factor $\partial_5\varepsilon\propto\left[\delta(y)-\delta(y-L)\right]$ in 
the step-function limit~\eqref{step_func} in Eq.~\eqref{radion_action}, while it should be 
positive on the both branes in the Goldberger-Wise setup. It would also be interesting to realize 
the Goldberger-Wise mechanism by introducing another bulk scalar field in our setup. 
%which can never be achieved in our setup 
%where $\partial_5\varepsilon\propto\left[\delta(y)-\delta(y-L)\right]$ in 
%the step-function limit~\eqref{step_func} in Eq.~\eqref{radion_action}. 
\item	The study including the higher order non-linear terms in the equation of motion for $\Phi$ would require a detailed numerical analysis, which we leave for future research.
Also, the back-reaction from $\Phi$ to the metric is neglected at this order. 
For a fixed four form field, we expect that gravitational instability would be absent 
after the stabilization of the radion. See also Refs.~\cite{Lehners:2005su,Maity:2006ub,Maity:2006in} for related issues.

%As is shown in the Eq.~(\label{radion_potential}), the radion potential indeed 
%becomes zero for $c_3=0$. It would be natural to expect that the radion potential 
%could be induced when we include the bulk quartic coupling $c_3\neq0$.
\end{itemize}

\section{Summary and Discussions}
We have proposed the modification of the five dimensional gravitational action that allows an arbitrary smoothing of the negative tension brane in the context of the Randall-Sundrum I brane world. This can be viewed not only as a possible regularization but also as arising from the high energy theory with the four form auxiliary gauge field.
It would be interesting to investigate possible connection of our mechanism to the four form mechanism in the five dimensional supergravity~\cite{Bergshoeff:2000zn,Fujita:2001bd}.

The application of our regularized $Z_2$-odd function $\varepsilon(y)$ to the $Z_2$-odd graviphoton gauge coupling might be possible in the context of the supersymmetric Randall-Sundrum model~\cite{Gherghetta:2000qt,Falkowski:2000er} with $\varepsilon(y)\varepsilon'(y)=0$ on the branes, while the version~\cite{Altendorfer:2000rr} 
with $\varepsilon(y)\varepsilon'(y)\neq0$ would require further elaboration.

We have discussed that the profile of the four form field can arise from 
the tachyon condensation in string theory.
In the point of view of tachyon condensation, it would be natural to expect the following 
scenario to be realized: 
The five-dimensional spacetime is filled with an unstable D-brane on which there exists a tachyon 
$T$. 
If $T$ forms a kink, then the kink would be regarded as a stable D3-brane to which $A_4$ couples. 
In other words, the braneworld is created dynamically. 
However, it seems strange that, in this scenario, an anit-kink corresponds to a negative-tension
brane since it is usually regarded as just an anti-brane with a positive tension. 
It would be worth studying this issue further. 

Our model does not allow the Goldberger-Wise mechanism for the radion stabilization. However we show that the introduction of the bulk scalar field and its natural embedding into our extended action can stabilize the radion. The stability under gravitational perturbation of the system would be worth investigating.

\bigskip

\section*{Acknowledgment}
We thank A.~Miwa, Y.~Sakamura and T.~Shiromizu for useful comments.
We acknowledge the YITP workshop YITP-W-06-11 on ``String Theory and Quantum Field Theory,'' Kyoto, Japan (2006), where partial results of this work is presented. 
This work was initiated in the Theoretical Physics Laboratory, RIKEN.
The research of N.Y.\ is supported in part by the JSPS Research Fellowships for Young Scientists.
The work of K.O.\ is partly supported by Scientific Grant by Ministry of Education and Science (Japan), Nos. 19740171, 20244028, and 20025004. 
The work of T.S. is supported in part by the Korea Research Foundation Leading Scientist Grant (R02-2004-000-10150-0), 
Star Faculty Grant (KRF-2005-084-C00003) and the Korea Research Foundation Grant, No. KRF-2007-314-C00056.

\end{document}